\documentclass{ws-procs975x65}

\def\beq{\begin{equation}}
\def\eeq{\end{equation}}

\begin{document}

\title{Conformal field theory of a space-filling string of gravitational ancestry}
\author{Claudio Bunster and Alfredo P\'{e}rez}

\address{Centro de Estudios Cient\'{i}ficos (CECs),\\ Av. Arturo Prat 514, Valdivia,
Chile\\
E-mail: bunster@cecs.cl, aperez@cecs.cl
}

\begin{abstract}
We present a classical conformal field theory on an arbitrary two-dimensional
spacetime background. The dynamical object is a space-filling string,
and the evolution may be thought as occurring on the manifold of the
conformal group. The theory is a ``descendant'' of the theory of
gravitation in two-dimensional spacetime. The discussion is based
on the relation of the deformations of the space-filling string with
conformal transformations. The realization of the conformal algebra
in terms of surface deformations possesses a classical central charge.
The action principle, the conformal and Weyl invariances of the action,
and the equations of motion are studied. The energy-momentum tensor,
the coupling to Liouville matter, and the cancellation of anomalies
are analyzed. The quantum theory is not discussed.
\end{abstract}


\bodymatter


\section{Introduction}

It took more than four decades since the publication of the Maxwell
equations to establish their conformal invariance. Thereafter the
conformal symmetry became a subject of continuous interest in theoretical
physics and the realm of its applications vast.\cite{Fradkin1,Fradkin2}

The conformal algebra in two-dimensional spacetime has the special
property of having an infinite number of generators. This makes the
symmetry especially powerful and makes conformal field theories in
two-dimensional spacetime appealing. Much work on them has been done
in recent years (see e.g. \refcite{Di-Francesco}, \refcite{Ginsparg}
and references therein).

The purpose of this report is to discuss a novel classical conformal
field theory, on an arbitrary two-dimensional Riemannian spacetime
background, in which the dynamical object is a space-filling string.

If one considers an arbitrary but fixed system of coordinates $y^{\lambda}$
on the background, a generic (parametrized) space-filling string is described
by a function $y^{\lambda}\left(x\right)$, and an infinitesimal deformation
is the change,
\begin{equation}
y^{\lambda}\left(x\right)\rightarrow y^{\lambda}\left(x\right)+\delta y^{\lambda}\left(x\right).\label{eq:yl}
\end{equation}

In two spacetime dimensions, and only then, that deformation may be
described in a conformally invariant manner as,
\begin{equation}
\delta y^{\lambda}=\delta\xi^{\perp}\tilde{n}^{\lambda}+\delta\xi^{1}y_{\;\;,1}^{\lambda},\label{eq:deltay}
\end{equation}
provided the normal $\tilde{n}^{\lambda}$ is chosen to be Weyl invariant,
instead of being the customary unit normal.

Then the deformation parameters $\delta\xi^{\perp}$, $\delta\xi^{1}$
describe a conformal transformation, because the general integrability
conditions of surface deformations\cite{Dirac,Schwinger,Teitelboim-Annals,Teitelboim-thesis}
imply that the corresponding generators $s_{\perp}$, $s_{1}$, obey
the algebra of the conformal group,\cite{Teitelboim-anomaly,Teitelboim-Two-dimensional} 

\begin{align}
\left[s_{\perp}\left(x\right),s_{\perp}\left(x'\right)\right] & =\left(s_{1}\left(x\right)+s_{1}\left(x'\right)\right)\delta'\left(x,x'\right),\label{eq:hphp-1-1}\\
\left[s_{1}\left(x\right),s_{\perp}\left(x'\right)\right] & =\left(s_{\perp}\left(x\right)+s_{\perp}\left(x'\right)\right)\delta'\left(x,x'\right)-\zeta\delta'''\left(x,x'\right),\label{eq:hphi-1-1}\\
\left[s_{1}\left(x\right),s_{1}\left(x'\right)\right] & =\left(s_{1}\left(x\right)+s_{1}\left(x'\right)\right)\delta'\left(x,x'\right),\label{eq:hihi-1-1}
\end{align}
where the central charge $\zeta$ is a constant with the units of
an action.%
\footnote{The conformal algebra (\ref{eq:hphp-1-1})--(\ref{eq:hihi-1-1}) can
be decomposed as two copies of the Virasoro algebra, $\left[L\left(x\right),L\left(x'\right)\right]=\left(L\left(x\right)+L\left(x'\right)\right)\delta'\left(x,x'\right)-\frac{\zeta}{2}\delta'''\left(x,x'\right)$,
whose generators are given by $L^{\pm}\left(x\right)=\left(1/2\right)\left(s_{\perp}\left(\pm x\right)\pm s_{1}\left(\pm x\right)\right)$.
With the standard convention the Virasoro central charge $c$ is given
by $c=12\pi\zeta$. %
} 

Therefore if one fixes a reference cut $y_{0}^{\lambda}\left(x\right)$
there is a one-to-one correspondence between the functions $y^{\lambda}\left(x\right)$
and the conformal transformation whose effect is to deform $y_{0}^{\lambda}\left(x\right)$
into $y^{\lambda}\left(x\right)$.\footnote{If one has a spacetime foliation $y^{\lambda}=y^{\lambda}(t,x)$,  then (\ref{eq:yl}) is equivalent to a change of the coordinates $t,x$. Therefore, a quantity, for example an action integral, will be  invariant under changes in the spacetime coordinates if and only if it is conformally invariant. For this reason we will use both terminologies interchangeably in the text, depending on the desired emphasis.}

We will discuss herein a realization of the algebra (\ref{eq:hphp-1-1})--(\ref{eq:hihi-1-1})
on a two-dimensional spacetime background, endowed with an arbitrary
but given metric $\gamma_{\lambda\rho}\left(y\right)$, in which \emph{the
conformal symmetry generators are built out of the $y^{\lambda}\left(x\right)$
themselves and their canonical conjugates}.

In virtue of the fact that the deformation is described in a conformally
invariant manner, one may say that the configuration space of the
theory is the group space of the conformal group.\textcolor{red}{}%
\footnote{The group space of the conformal group and the related concept of
exponentiation of an element of the conformal Lie algebra are not
devoid of subtleties, for a lucid discussion see e.g. Ref. \refcite{Azcarraga}.%
} 

The action and symmetry generators of the theory in question were
found in Ref. \refcite{BP}, as those of a particular---and
especial---``G-brane.'' The purpose of the present discussion is
to bring out in a self-contained manner the (1+1) G-brane as a classical
conformal field theory, focusing on aspects that were not covered
or emphasized in Ref. \refcite{BP}.

The plan of the paper is the following: Sec. \ref{sec:Surface-deformation-algebra}
recalls the results obtained in Ref. \refcite{BP} for the
conformal generators of the (1+1) G-brane, while Sec. \ref{sec:Action-principle}
discusses the action principle, the (lack of) conformal and Weyl invariance
of the action, and the equations of motion. Next Sec. \ref{sec:Stress-tensor.-Coupling}
is devoted to analyzing the energy-momentum tensor, the coupling to
Liouville matter, and the cancellation of anomalies. Finally Sec.
\ref{sec:Concluding-remarks} contains concluding remarks.

\section{Surface deformation algebra and conformal symmetry\label{sec:Surface-deformation-algebra}}

\subsection{Weyl invariant normal}

To relate the infinitesimal deformation (\ref{eq:yl}) to the conformal
algebra, it is essential to bring in the metric $\gamma_{\lambda\rho}$
in order to split $\delta y^{\lambda}$ into its normal and tangential
components, $\delta\xi^{\perp}$ and $\delta\xi^{1}$ according to
eq. (\ref{eq:deltay}). The normal $\tilde{n}^{\lambda}$ appearing
in that equation is defined by, 
\begin{equation}
\gamma_{\lambda\rho}\tilde{n}^{\lambda}\tilde{n}^{\rho}=-g_{11}\equiv-g,\label{eq:normal}
\end{equation}
where,

\begin{equation}
g_{11}=\gamma_{\lambda\rho}y_{\;\;,1}^{\lambda}y_{\;\;,1}^{\rho},\label{eq:indmetric}
\end{equation}
is the metric on the cut $y^{\lambda}\left(x\right)$. On account
of (\ref{eq:normal}) and (\ref{eq:indmetric}) the normal $\tilde{n}^{\lambda}$
is invariant under Weyl transformations,
\begin{equation}
\gamma_{\lambda\rho}\left(y\right)\rightarrow e^{\sigma\left(y\right)}\gamma_{\lambda\rho}\left(y\right).\label{eq:Weyl}
\end{equation}

The fields $y^{\lambda}$, $y_{\;\;,1}^{\lambda}$, and $w^{\lambda}$
are also Weyl invariant because the background metric $\gamma_{\lambda\rho}$
does not enter into their definition.

\subsection{Canonical generators}

We now recall the expressions for the generators $s_{\perp}$ and
$s_{1}$ obtained in Ref. \refcite{BP}.

If the canonical momenta conjugate to $y^{\lambda}$ are denoted by
$w_{\lambda}$, so that in terms of Poisson brackets,
\[
\left[y^{\lambda}\left(x\right),w_{\text{\ensuremath{\rho}}}\left(x'\right)\right]=\delta_{\rho}^{\lambda}\delta\left(x,x'\right),
\]
the canonical generators that span the algebra (\ref{eq:hphp-1-1})--(\ref{eq:hihi-1-1})
are realized in terms of the fields $y^{\lambda},w_{\rho}$ as,
\begin{align*}
s_{\perp} & =w_{\perp}-\mathcal{L},\\
s_{1} & =w_{1}.
\end{align*}
Here, 
\begin{align*}
w_{\perp} & =w_{\lambda}\tilde{n}^{\lambda},\\
w_{1} & =w_{\lambda}y_{\;\;,1}^{\lambda},
\end{align*}
and, 
\begin{equation}
\mathcal{L}=\frac{\zeta}{4}\left(2e^{\varphi}K^{2}-\frac{1}{2}\varphi'^{2}+2\varphi''+\Lambda e^{\varphi}\right).\label{eq:L}
\end{equation}
The fields $\varphi$ and $K$ are the local relative scale and extrinsic
curvature of the space-filling string respectively. They are expressed
in terms of the $y^{\lambda}$as, 
\begin{equation}
\varphi=\log\left(g_{11}\right)=\log\left(\gamma_{\lambda\rho}y_{\;\;,1}^{\lambda}y_{\;\;,1}^{\rho}\right),\label{eq:phi}
\end{equation}
\begin{equation}
K=g^{-3/2}\tilde{n}_{\lambda}D_{1}y_{\;\;,1}^{\lambda}\,.\label{eq:K}
\end{equation}
The derivative $D_{1}=y_{\;\;,1}^{\lambda}D_{\lambda}$ is the covariant
derivative in the external space projected onto the string.

If in (\ref{eq:L}) the field $\varphi$ is regarded as a fundamental
variable, and $K$ as its ``invariant velocity''; as opposed to
being expressed in terms of the $y^{\lambda}$ through (\ref{eq:phi})
and (\ref{eq:K}), then $\mathcal{L}$ appearing in (\ref{eq:L})
becomes the Lagrangian for an analog in two-dimensional spacetime
of Einstein's theory of gravitation \refcite{Teitelboim-anomaly},
\refcite{Teitelboim-Two-dimensional}, \refcite{Jackiw}, which is related
to the Liouville theory. If the replacements (\ref{eq:phi}) and (\ref{eq:K})
are implemented the ``gravitational ancestor'' (\ref{eq:L}) gives
rise to its descendant, the space-filling string.

\section{Action principle\label{sec:Action-principle}}

The action integral is given by,
\begin{equation}
I=\int dtdx\left(w_{\lambda}\dot{y}^{\lambda}-\eta^{\perp}s_{\perp}-\eta^{1}s_{1}\right).\label{eq:action}
\end{equation}

The Weyl invariant lapse and shift functions $\eta^{\perp},\eta^{1}$
are considered as external fields which are not varied in the action
principle. They parametrize the most general conformal transformation
whose unfolding in time is the dynamics of the system, generated by
the Hamiltonian, 
\[
H=\int dx\left(\eta^{\perp}s_{\perp}+\eta^{1}s_{1}\right).
\]
A particular case is, $\eta^{\perp}=1$, $\eta^{1}=0$, which might
be termed a ``rigid time translation'', the corresponding Hamiltonian
is the integral of $s_{\perp}$.

\subsection{Hamiltonian linear in the momenta.}

The Hamiltonian is linear in the momenta because it is a linear combination
of the generators $s_{\perp},$ $s_{1}$, which have that property. This is a distinctive feature of the
(1+1) G-brane. Geometrically, it captures the fact that the motion may be considered
as a displacement on the manifold of the conformal group (see eq.
(\ref{eq:EOM1}) below). Dynamically it implies that, independently
of the sign of the constant $\zeta$, the ``energy density'' $s_{\perp}$
can be arbitrarily large and positive or negative.
As a consequence of the  linearity  in the momenta they cannot be expressed  in terms of the velocities
to obtain an action principle in terms of a Lagrangian which would
depend on $y$ and $\dot{y}$. This situation changes however when one couples the (1+1) G-brane to matter as discussed in section \ref{Lag}.

\subsection{Change of the action under a conformal transformation\label{sub:Change-of-the}}

It will be useful here to employ a standard compact notation to account
for both continuous and discrete indices, so that, for example summation
over $a'$ includes an integration over an accompanying continuous
index $x'$. The conformal algebra (\ref{eq:hphp-1-1})--(\ref{eq:hihi-1-1})
then reads,
\[
\left[s_{a},s_{b'}\right]=\kappa_{ab'}^{\;\;\; c''}s_{c''}+\zeta_{ab'},
\]
where the structure constants $\kappa$ are products of $\delta$-functions
and the central term $\zeta$ is proportional to $\delta'''$.

The conformal transformation of any functional $F$ of the dynamical
fields $y^{\lambda}$ and $w_{\lambda}$ is given by its Poisson bracket
with the corresponding generator:
\begin{equation}
\delta_{\xi}F=\left[F,\xi^{a}s_{a}\right],\label{eq:deltaF}
\end{equation}
whereas the change of the ``conformal lapse'' and shift $\eta^{a}=\left(\eta^{\perp},\eta^{1}\right)$
is given by,
\begin{equation}
\delta\eta^{a''}=\dot{\xi}^{a''}-\kappa_{bc'}^{\;\;\; a''}\eta^{b}\xi^{c'}.\label{eq:deltaeta}
\end{equation}

If one employs equations (\ref{eq:deltaF}) and (\ref{eq:deltaeta})
to evaluate the change in the action (\ref{eq:action}) for a region
$M$ with boundary $\partial M$, one finds

\begin{align}
\delta I & =\int_{\partial M}\xi^{\perp}\mathcal{L}+\int_{M}\xi^{a}\eta^{b'}\zeta_{ab'},\nonumber \\
 & =\int_{\partial M}\xi^{\perp}\mathcal{L}+\zeta\int_{M}\left(\eta^{\perp}\xi^{1'''}-\xi^{\perp}\eta^{1'''}\right).\label{eq:lack}
\end{align}
Therefore, due to the central term, even when the boundary is mapped
onto itself ($\xi^{\perp}=0$ over $\partial M$), the action is not
conformally invariant. This anomaly does not affect the equations
of motion because the central term does not include the dynamical
fields.

\subsection{Change of the action under a Weyl transformation}

Under an infinitesimal Weyl transformations of the background metric,
\[
\delta\gamma_{\lambda\rho}=\sigma\gamma_{\lambda\rho},
\]
the action (\ref{eq:action}) transform as:
\begin{equation}
\delta I=-\frac{\zeta}{4}\int dtdx\eta^{\perp}e^{\varphi}\left(R-\Lambda\right)\sigma.\label{eq:deltaweyl}
\end{equation}
The change (\ref{eq:deltaweyl}) may be obtained by direct calculation,
or, better, by realizing that it must be the same as the one of its
gravitational ancestor mentioned at the end of the previous section.
In that context eq. (\ref{eq:deltaweyl}) is at the heart of the gravitation
theory in two spacetime dimensions: demanding that it should vanish
gives then the equation of motion,
\[
R-\Lambda=0,
\]
of the ancestor theory.\cite{Teitelboim-anomaly,Teitelboim-Two-dimensional,Jackiw}

\subsection{Equations of motion}

For general $\eta^{\perp}$ and $\eta^{1}$ the equations of motion
are given by,
\begin{eqnarray}
\dot{y}^{\lambda} & = & \eta^{\perp}\tilde{n}^{\lambda}+\eta^{1}y_{\;\;,1}^{\lambda},\label{eq:EOM1}\\
\dot{w}_{\perp} & = & 2w_{1}\eta^{\perp'}+\eta^{\perp}w_{1}'+2\eta^{1'}\left(w_{\perp}-\mathcal{L}\right)+\eta^{1}\left(w_{\perp}'-\mathcal{L}'\right)-\zeta\eta^{1'''}+\dot{\mathcal{L}},\label{eq:EOM2}\\
\dot{w}_{1} & = & 2\left(w_{\perp}-\mathcal{L}\right)\eta^{\perp'}+\eta^{\perp}\left(w_{\perp}'-\mathcal{L}'\right)+2w_{1}\eta^{1'}+\eta^{1}w_{1}'-\zeta\eta^{\perp'''}.\label{eq:EOM3}
\end{eqnarray}
These equations are conformally invariant because, as it was stated
at the end of subsection \ref{sub:Change-of-the} the change in the
action under a conformal transformation does not depend on the dynamical
fields. Comparison of equation (\ref{eq:EOM1}) with equation (\ref{eq:deltay})
shows that the motion may be considered as a displacement on the manifold
of the conformal group. 

To obtain insight we now consider the case in which the background
is Minkowski space and the external coordinate system is cartesian,
that is 
\begin{align*}
\gamma_{\alpha\beta} & =\eta_{\alpha\beta}.
\end{align*}
Furthermore we take at time $t=0$
\begin{equation}
y^{0}=0,\quad\quad y^{1}=x,\label{eq:initial}
\end{equation}
while leaving the $w_{\lambda}\left(t=0,x\right)$ arbitrary, and
further specialize to a ``rigid time translation'', i.e.,
\[
\eta^{\perp}=1,\quad\quad\eta^{1}=0,
\]
for all times. 

Then the solution of the equations of motion is:
\begin{align}
y^{0} & =t,\nonumber \\
y^{1} & =x,\nonumber \\
w_{\perp} & =w_{+}\left(x+t\right)+w_{-}\left(x-t\right),\label{eq:solution}\\
w_{1} & =w_{+}\left(x+t\right)-w_{-}\left(x-t\right).\nonumber 
\end{align}
It is important to emphasize that the above solution \emph{is not}
the most general solution of the equations of motion when $\gamma_{\alpha\beta}=\eta_{\alpha\beta}$
because the initial conditions (\ref{eq:initial}) are very particular
corresponding to a simple cut through the spacetime. The most general
solution is obtained by acting on (\ref{eq:solution}) with the conformal
group through an iteration of transformations generated by $s_{\perp}$
and $s_{1}$. The result of this iteration is to go from the cut (\ref{eq:initial})
to a generic spacelike cut $y^{\lambda}\left(t,x\right)$, but the
corresponding result for $w_{\lambda}$ cannot written in a closed
form. It is important to keep in mind that different cuts $y^{\lambda}\left(t,x\right)$
do not correspond to different gauge choices, because $s_{\perp}$
and $s_{1}$ do not generate a gauge symmetry since they are not constrained
to vanish. Note that for the solution (\ref{eq:solution}) the conformal
energy density $s_{\perp}$ does not vanish when $w_{\lambda}=0$,
but is given by the background energy, 
\[
-\frac{\Lambda\zeta}{4}.
\]

\section{Energy-momentum tensor. Coupling to matter. Anomaly cancellation\label{sec:Stress-tensor.-Coupling}}

\subsection{Energy-momentum tensor}

One may define an energy-momentum tensor for the (1+1) G-brane through
the standard formula,
\begin{equation}
T_{\lambda\rho}=-\frac{2}{\sqrt{-\gamma}}\frac{\delta I}{\delta\gamma^{\lambda\rho}}.\label{eq:T}
\end{equation}
As a consequence of the lack of invariance of the action under changes
of the spacetime coordinates expressed by (\ref{eq:lack}), $T_{\lambda\rho}$
defined by (\ref{eq:T}) not only is not conserved when the equations
of motion (\ref{eq:EOM1})--(\ref{eq:EOM3}) hold but its components
do not even transform as those of a tensor under a change of the spacetime
coordinates. This conformal anomaly is brought in by the presence
of the central charge (``Schwinger term'') in the commutation rule
for the energy and momentum densities $s_{\perp}$, $s_{1}$, which
are related to $T_{\lambda\rho}$ by, 
\begin{align}
s_{\perp} & =T_{\lambda\rho}\tilde{n}^{\lambda}\tilde{n}^{\rho},\label{eq:sp}\\
s_{1} & =T_{\lambda\rho}\tilde{n}^{\lambda}y_{\;\;,1}^{\rho}.\label{eq:s1}
\end{align}

The remaining component, 
\[
T_{11}=T_{\lambda\rho}y_{\;\;,1}^{\lambda}y_{\;\;,1}^{\rho}=gT+s_{\perp},
\]
where, 
\[
T=\gamma^{\lambda\rho}T_{\lambda\rho},
\]
may be obtained from (\ref{eq:deltaweyl}) to be,
\begin{equation}
T=\frac{\zeta}{2}\left(R-\Lambda\right),\label{tracestress}
\end{equation}
and it is a world scalar although $T_{\lambda\rho}$ is not a tensor.

\subsection{Coupling to Liouville field}

One may couple the (1+1) G-brane by adding to (\ref{eq:action}) the
action of a ``matter field'' also described by a conformal field
theory. That matter action will have a form similar to (\ref{eq:action}):
\[
I_{\text{matter}}=\int\pi\dot{\psi}-\eta^{\perp}s_{\perp}^{\text{matter}}-\eta^{1}s_{1}^{\text{matter}}.
\]
The generators $s^{\text{Liouv}}$ will obey the conformal algebra
(\ref{eq:hphp-1-1})--(\ref{eq:hihi-1-1}). Since they will be built
out of $\pi$ and $\psi$ they will commute with those of the (1+1)
G-brane. Therefore the generators of the complete theory,
\[
s^{\text{total}}=s+s^{\text{\text{matter}}},
\]
will obey the conformal algebra with a total central charge which
is the sum of the central charges of the (1+1) G-brane and that of
the matter theory.

There is one theory for a matter field which has a central charge
at the classical level, that is the Liouville theory for which, 
\begin{align*}
s_{\perp}^{\text{Liouv}} & =\frac{1}{2}\left(k\pi^{2}+k^{-1}\psi'^{2}\right)-2k^{-1}\psi''+\frac{1}{2k}m^{2}e^{\psi},\\
s_{1}^{\text{Liouv}} & =\pi\psi'-2\pi',
\end{align*}
obey the algebra (\ref{eq:hphp-1-1})--(\ref{eq:hihi-1-1}) with,
\[
\zeta^{\text{Liouv}}=\frac{4}{k}.
\]
 The total central charge is then
\begin{equation}
\zeta^{\text{total}}=\zeta+\frac{4}{k}.\label{eq:sum}
\end{equation}

\subsection{Anomaly cancellation}

An interesting possibility now appears, namely that of adjusting the
(1+1) G-brane central charge $\zeta$ for a given $k$ so that the
sum (\ref{eq:sum}) vanishes,
\begin{equation}
\zeta^{\text{total}}=0.\label{eq:zetazero}
\end{equation}
For physical reasons, such as the positivity of the energy of the
Liouville field, it is desirable to have $k>0$, however no similar
restriction appears for $\zeta$ because, as it was pointed out above,
for any sign of $\zeta$ the (1+1) G-brane energy is unbounded from
above and from below.

When (\ref{eq:zetazero}) holds, and only then, one has the option
of constraining the total generators to vanish,
\begin{align}
s_{\perp}^{\text{total}} & =s_{\perp}+s_{\perp}^{\text{\text{Liouv}}}\approx0,\label{eq:sptot}\\
s_{1}^{\text{total}} & =s_{1}+s_{1}^{\text{\text{Liouv}}}\approx0.\label{eq:s1tot}
\end{align}
If the total central charge were not to vanish equations (\ref{eq:sptot})
and (\ref{eq:s1tot}) would not be preserved under the action of $s_{\perp}^{\text{total}}$
and $s_{1}^{\text{total}}$.

Before the constraints are imposed the complete theory is invariant under two
independent conformal symmetries, one acting on the (1+1) G-brane,
the other on the Liouville field. There are independent parameters $\eta$ for each, which are not varied in the action principle. The
two theories are decoupled and there are three degrees of freedom
per space point.

If the constraints are imposed the only symmetry is a simultaneous conformal transformation of the same magnitude on both the brane and the Liouville field. The common parameter $\eta$
is now varied in the action principle in order to yield the constraint. The number of degrees of freedom per point of the
combined theory then decreases from three to one, and the brane and
the Liouville field become coupled through the constraint, although
their equations of motion remain uncoupled.

When the total central charge vanishes, according to eq. (\ref{eq:zetazero}) the complete action is invariant under spacetime reparametrizations because the anomalous term in (\ref{eq:lack}) is proportional to $\zeta^{\text{total}}$. On the other hand the change of the complete action under a Weyl transformation is still given by (\ref{eq:deltaweyl}) because the Liouville action is Weyl invariant if one assumes $\psi$ to have that property.

\subsection{Lagrangian\label{Lag}}

For the coupled action one may pass from the Hamiltonian formulation to a Lagrangian formulation by eliminating the constraints and the momenta from their equations of motion. The resulting Lagrangian density which is a function of $y^{\lambda}$, $\dot{y}^{\lambda}$, $\psi$, $\dot{\psi}$ and $\gamma_{\lambda \rho}$ has the form,
\[
\mathcal{L}^{\text{total}}=\mathcal{L}^{\text{string}}+\mathcal{L}^{\text{Liouv}},
\]
where,
\[
\mathcal{L}^{\text{string}}=\eta^{\perp}\mathcal{L},
\]
with $\mathcal{L}$ given by (\ref{eq:L}), and 
\begin{equation*}
\mathcal{L}^{\text{Liouv}}=\frac{1}{2k}\left\{ \left(\eta^{\perp}\right)^{-1}\left[\dot{\psi}-\psi'\eta^{1}-2\left(\eta^{1}\right)'\right]^{2}-\eta^{\perp}\psi'^{2}+4\eta^{\perp}\psi''+2\eta^{\perp}\Lambda e^{\psi}\right\}. 
\end{equation*}
(See Ref. \refcite{Teitelboim-anomaly}).

 Here $\eta^{\perp}$ and $\eta^{1}$ stand as abbreviations for,\[
\eta^{\perp}=-g^{-1}\tilde{n}_{\lambda}\dot{y}^{\lambda},\quad\quad\eta^{1}=g^{-1}\gamma_{\lambda\rho}y_{\:\:,1}^{\lambda}\dot{y}^{\rho},
\]
and the constant $k$ is related to the (1+1) G-brane central charge $\zeta$ through (\ref{eq:sum}) and (\ref{eq:zetazero}).

Since the conformal anomalies of the (1+1) G-brane and the Liouville field mutually cancel the action,\[
I^{\text{total}}=\int dtdx\mathcal{L}^{\text{total}},
\]
is invariant under spacetime reparametrizations, therefore the total energy-momentum tensor, \begin{equation*}
T^{\text{total}}_{\lambda\rho}=-\frac{2}{\sqrt{-\gamma}}\frac{\delta I^{\text{total}}}{\delta\gamma^{\lambda\rho}},
\end{equation*}
is indeed a tensor and it is conserved. The energy and momentum densities (of weight two) are given by \begin{eqnarray*}
T_{\perp\perp} & = & T_{\lambda\rho}^{\text{total}}\tilde{n}^{\lambda}\tilde{n}^{\rho}=T_{\perp\perp}^{\text{Liouv}}-\mathcal{L},\\
T_{\perp1} & = & T_{\lambda\rho}^{\text{total}}\tilde{n}^{\lambda}y_{\:\:,1}^{\rho}=T_{\perp1}^{\text{Liouv}},
\end{eqnarray*}
whereas the trace is still given by (\ref{tracestress}).

One sees that the energy of the system acquires a ``background contribution'' $-\mathcal{L}$ in addition to the Liouville energy density. Furthemore it is only the sum of the two which has covariant meaning. If one changes the spatial coordinate on the surface $y^{\lambda}(x)$, the two pieces do not transform separately as a density because of the conformal anomaly, but the sum does.

In the above discussion the Liouville field $\psi$ is unrelated to
the field $\varphi$ appearing in the ancestor gravitation theory
referred to after eq. (\ref{eq:K}). If it where identified with it,
so as to couple the descendant to its ancestor, the variation of the
combined action with respect to $\psi$ would be,
\[
\delta I=-\int dtdxk^{-1}\eta^{\perp}e^{\psi}\left(\Lambda+\frac{m^{2}}{2}\right)\delta\psi,
\]
so that the action principle would either be contradictory, if $\Lambda\neq-m^{2}/2$, 
or empty, if $\Lambda=-m^{2}/2$. Incest is either forbidden or it
does not happen.

\section{Concluding remarks\label{sec:Concluding-remarks}}

We have presented a conformal field theory on an arbitrary two-dimensional
spacetime background. The dynamical object is a space-filling string,
and the evolution may be thought as occurring on the group space of
the conformal group. The theory is a ``descendant'' in the sense
of Ref. \refcite{BP}, of an analog of the theory of gravitation
proposed in Refs. \refcite{Teitelboim-anomaly}, \refcite{Teitelboim-Two-dimensional}
for two dimensional spacetime. The discussion has remained classical
throughout, including the treatment of the central charge which already
appears at that level. One would expect many delicate issues to appear
in the passage to the quantum theory, not least in the discussion
of anomalies.
\section*{Acknowledgments}

The Centro de Estudios Cient\'{i}ficos (CECs) is funded by the Chilean
Government through the Centers of Excellence Base Financing Program
of Conicyt. C.B. wishes to thank the Alexander von Humboldt Foundation
for a Humboldt Research Award. The work of A.P. is partially funded
by the Fondecyt Grant Nº 11130262.

\end{document}